\documentclass[twocolumn,showpacs,preprintnumbers,amsmath,amssymb]{revtex4}
\usepackage{graphicx}
\begin{document}
\preprint{IMAFF-RCA-04-05}
\title{Achronal cosmic future}

\author{Pedro F. Gonz\'{a}lez-D\'{\i}az}
\affiliation{Colina de los Chopos, Centro de F\'{\i}sica ``Miguel A.
Catal\'{a}n'', Instituto de Matem\'{a}ticas y F\'{\i}sica Fundamental,\\
Consejo Superior de Investigaciones Cient\'{\i}ficas, Serrano 121,
28006 Madrid (SPAIN).}
\date{\today}
\begin{abstract}
The spherically symmetric accretion of dark and phantom energy
onto Morris-Thorne wormholes is considered. It is obtained that
the accretion of phantom energy leads to a gradual increase of the
wormhole throat radius which eventually overtakes the
super-accelerated expansion of the universe and becomes infinite
at a time in the future before the occurrence of the big rip
singularity. After that time, as it continues accreting phantom
energy, the wormhole becomes an Einstein-Rosen bridge whose
corresponding mass decreases rapidly and vanishes at the big rip.
\end{abstract}


\maketitle

The universe which we live in is subject to a current evolution
which is certainly causal. That happens no matter whether it is
expanding in an accelerating fashion or not. But one cannot be so
sure about causality in the primordial universe. A rather natural
scenario was in fact proposed by Gott [1] in which the universe
created itself in a classically and quantum-mechanically stable
way [2], starting with a noncausal patch filled with closed
timelike curves which might be followed by an immediate ekpyrotic
quantum evolution [3]. However, nobody keeps any doubts about
causality in the future. There is a rather tacit consensus that,
whatever the fate of the universe, it will be governed by causal,
chronal principles which probably respect the second law of
thermodynamics [4]. It is currently thought following present
scientific logic that, after all, the present universe is big
enough as for consistently admitting a classical description where
no time traveling or similar oddities that disrupt causal
evolution appear to be allowed [5]. We are going nevertheless to
consider a possible cosmological framework where really there
could be a disruption of the causal evolution of the accelerating
universe in the future. Actually such a prediction appears to be
unavoidable if the universal equation of state for dark energy is
characterized by a constant parameter $w=p/\rho <-1$ (where $p$
and $\rho$ respectively denote pressure and energy density), a
case which is dubbed phantom energy [6] and which is not at all
excluded by present cosmological constraints [7]. That phantom
energy has a fair collection of weird properties which are all
based on the requirement that the phantom stuff violates the
dominant energy condition, that is $p+\rho<0$ [8]. But violating
that energy condition would also make it possible the existence of
natural traversible wormholes [9], even at scales larger than the
Planck scale, if were it not for the fact that such classical
wormholes appear to be unstable to quantum vacuum fluctuations
[5]. The evolution of wormholes and ringholes [10] induced by the
accelerated expansion of the universe has been already considered
[11] for the case of dark energy with both $w>-1$ and $w<-1$. It
was seen that in all cases the scale of the tunneling increased as
the scale factor becomes larger, and blew up at the big rip [11]
when the universe was filled with phantom energy. Once we have
learnt what kind of kinematic effects can be expected from dark
energy on wormholes, we shall consider in this letter the effect
of dark energy accretion onto submicroscopic and macroscopic
wormholes. Our main result is that, as a consequence of dark
energy accretion, whereas the wormhole throat gradually decreases
down to a minimum size if $w>-1$, whenever $w<-1$ the wormhole
throat grows rapidly up to reaching an infinite size before the
universe gets into the big rip singularity.

The mass $\mu$ of the spherical thin shell of exotic matter in a
Morris-Thorne wormhole can be given by [12,13]
\begin{equation}
\mu=-\pi b_0/2 ,
\end{equation}
where we have used units such that $G=\hbar=c=1$, and $b_0$ is the
radius of the spherical wormhole throat. Now, since, very
approximately, the mass in Eq. (1) is just the negative of the
amount of mass required to produce a normal Schwarzschild wormhole
(i.e. the wormhole connecting a black hole with mass $M\simeq-\mu$
to its corresponding white hole, making an Einstein-Rosen bridge
[14]), the rate at which the exotic mass of a wormhole changes by
accreation of dark energy can be approximately equalized to the
negative of the rate of the mass change of the black hole making
the Schwarzschild wormhole, due to accreation of dark energy.
Quite recently, Babichev, Dokuchaev and Eroshenko [15] have
obtained that, as a consequence of fluid accretion, the mass of a
black hole changes at a rate $\dot{M}=4\pi AM^2(p+\rho)$, where
$A$ is a positive dimensionless constant. For an equation of state
of the fluid $p=w\rho$, which will be assumed to be constant
throughout the present letter, we can therefore write for the rate
of change the throat radius of a Morris-Thorne wormhole due to
dark energy accretion the expression
\begin{equation}
\dot{b}_0=-2\pi^2 Db_0^2(1+w)\rho,
\end{equation}
with $D$ another positive dimensionless constant, $D\simeq A$, and
$\rho$ the energy density of the dark energy fluid.

It is known that the dark energy density is given by
\begin{equation}
\rho=\rho_0 a^{-3(1+w)} ,
\end{equation}
where $\rho_0$ is an arbitrary constant playing the role of the
initial value of the energy density at the onset of dark energy
domination, and $a=a(t)$ is the scale factor. A general solution
for that scale factor in the case of a universe dominated by dark
energy can be written as [11]
\begin{eqnarray}
&&a(t)=\left[a_0^{3(1+w)/2}
+\frac{3}{2}(1+w)\sqrt{\frac{8\pi\rho_0}{3}}(t-
t_0)\right]^{2/[3(1+w)]}\nonumber\\ &&\equiv T^{2/[3(1+w)]} ,
\end{eqnarray}
in which $a_0$ and $t_0$ respectively are the initial values for
the scale factor and time at the onset of dark energy domination.
From Eqs. (1), (2) and (3) we have then
\begin{equation}
\dot{b}_0= -2\pi^2 D(1+w)\rho_0 b_0^2 T^{-2} .
\end{equation}
Trivial integration of this equation finally produces
\begin{equation}
b_0=\frac{b_{0i}}{1+\frac{b_{0i}(t-t_0)}{\dot{b}_{0i}a_0^{3(1+
w)/2}T}} ,
\end{equation}
\begin{equation}
\dot{b}_{0i}=\frac{1}{2\pi^2 D\rho_0(1+w)} .
\end{equation}

Thus, for all quintessential models with $w>-1$, the radius of the
wormhole throat, and hence the mass of exotic matter contained in
it, will gradually decrease with time and tends to a constant
minimum value
\begin{equation}
b_{0{\rm min}}=\frac{b_{0i}}{1+ \frac{4\pi^2 D\rho_0
b_{0i}}{3a_0^{3(1+w)/2}\sqrt{8\pi\rho_0/3}}} ,
\end{equation}
as $t\rightarrow\infty$. The case of a universe filled with
phantom energy for which $w<-1$ is actually quite more
interesting. If $w<-1$ we have in fact
\begin{equation}
b_0=\frac{b_{0i}}{1-\frac{(t-t_0)b_{0i}}{\dot{b}_{0i}'(t_*-
t_0)T'}} ,
\end{equation}
\begin{equation}
\dot{b}_{0i}'=\frac{3}{4\pi^2 D\rho_0} ,
\end{equation}
\begin{equation}
T'=a_0^{-3(|w|-1)/2}-\frac{3}{2}(|w|- 1)\sqrt{8\pi\rho_0/3}(t-t_0)
,
\end{equation}
\begin{equation}
t_*=t_0+\frac{2}{3(|w|-1)a_0^{3(|w|- 1)/2}\sqrt{8\pi\rho_0/3}} ,
\end{equation}
$t_*$ being the time at which the big rip singularity takes place.
While the remarkable, but not fully unexpected, prediction of Eq.
(9) is that when the wormhole accretes phantom energy the radius
of its throat will gradually increase (notice that after all, the
wormhole contains the same kind of phantom energy as the stuff
that it is accreting), that equation also shows the rather
surprising result that, as a consequence from such an exotic
matter increase, the size of the spherical wormhole throat will
turn out to blow up at a time
\begin{equation}
\tilde{t}=t_0 +\frac{t_*-t_0}{1+3\dot{b}_{0i}'
b_{0i}a_0^{3(|w|-1)/2}}
\end{equation}
which necessarily occurs before the big rip singularity at $t_*$
(see Fig. 1). We note however that at time $t=\tilde{t}$ the
exotic energy density becomes zero, and that there is not any
curvature singularity [9]; therefore one needs not cutting the
evolution of the universe at that point. Notice as well that after
$\tilde{t}$ the mass of the wormhole, $\mu$, becomes positive, and
this means that what initially was a Morris-Thorne wormhole is
converted after time $\tilde{t}$ into an Einstein-Rosen bridge
which will immediately pinches off to leave a black-white hole
pair that will rapidly loss its mass by the
Babichev-Dokuchaev-Eroshenko mechanism [15] to vanish at the big
rip.

\begin{figure}
\includegraphics[width=.9\columnwidth]{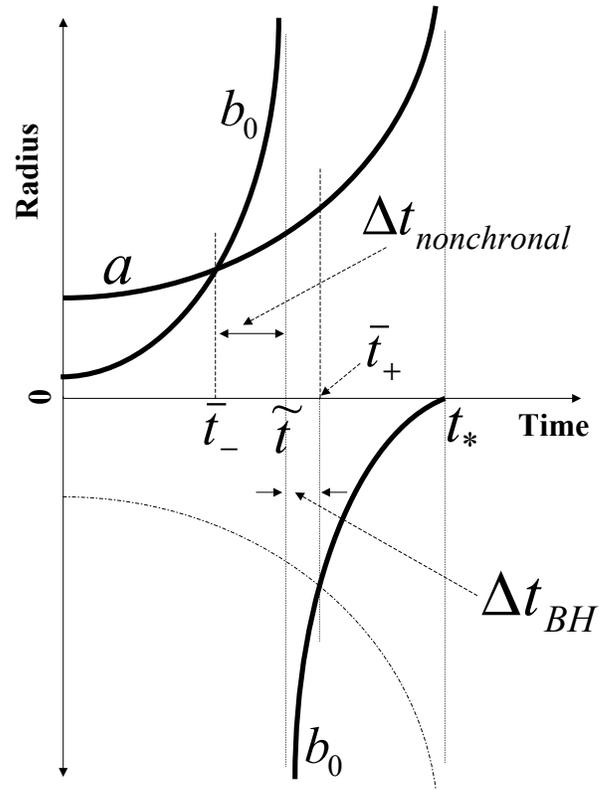}
\caption{\label{fig:epsart} Evolution of the radius of the
wormhole throat, $b_0$, induced by accretion of phantom energy. At
time $t=\tilde{t}$, the negative exotic mass becomes infinite and
then changes sign, so converting the wormhole into an
Einstein-Rosen bridge whose associated mass decreases down to zero
at the big rip at $t=t_*$. During the time interval $\Delta
t_{nonchronal}$ there will be a disruption of the causal evolution
of the whole universe.}
\end{figure}

The above results can be expected to happen in all dark energy
models that predict a big rip singularity in the future once they
are enforced to violate the dominant energy condition. Thus, in
the essentially distinct case where the dark energy which is
accreted by the Morris-Thorne wormhole corresponds to what is
denoted as a k-essence field [16], equipped with non-canonical
kinetic energy, we have for the phantom regime [17]
\begin{equation}
p+\rho=-3(1-\xi)H^2/\xi ,
\end{equation}
with $0<\xi<1$ and $H=\dot{a}/a$, where
\begin{equation}
a\propto (t-t_b)^{-2\xi/[3(1-\xi)]} ,
\end{equation}
in which $t_b$ is an arbitrary time at which the big rip takes
place, we obtain
\begin{equation}
b_0=\frac{b_{0i}}{1-\frac{tb_{0i}}{\dot{b}_{0i}t_b(t_b-t)}},
\end{equation}
\begin{equation}
\dot{b}_{0i}=\frac{3(1-\xi)}{8\pi^2 D\xi} .
\end{equation}
It can be readily seen that the properties of Eq. (16)
qualitatively match those implied by Eq. (9), with the time at
which the radius of the wormhole throat becomes infinite being
given in this case by
\begin{equation}
\tilde{t}=\frac{t_b}{1+\frac{b_{0i}}{\dot{b}_{0i}t_b}} ,
\end{equation}
i.e. again $\tilde{t}<t_b$. Exactly the same expressions as those
given in Eqs. (16) and (18) (but for a slightly different
$\dot{b}_{0i}$) are also obtained when one uses the simple
solution employed by Babichev, Dokuchaev and Eroshenko [15].

\begin{figure}
\includegraphics[width=.9\columnwidth]{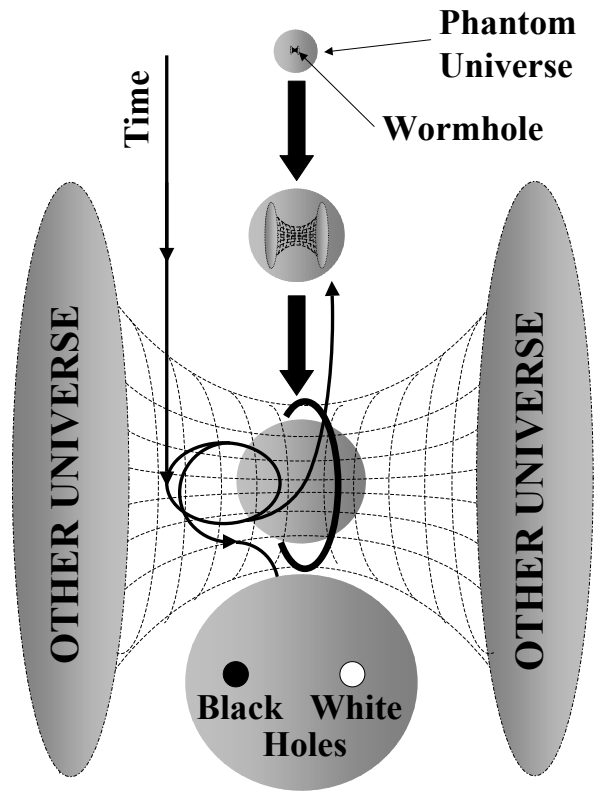}
\caption{\label{fig:epsart} Pictorial representation of the
evolution of a universe filled with phantom energy which is being
accreted onto an initially sub-microscopic wormhole for the case
(A) (see the text) when the wormhole is re-connected to other two
extra larger universes, during the time period on which the
wormhole throat is larger than the universe and time follows
closed curves.}
\end{figure}

\begin{figure}
\includegraphics[width=.9\columnwidth]{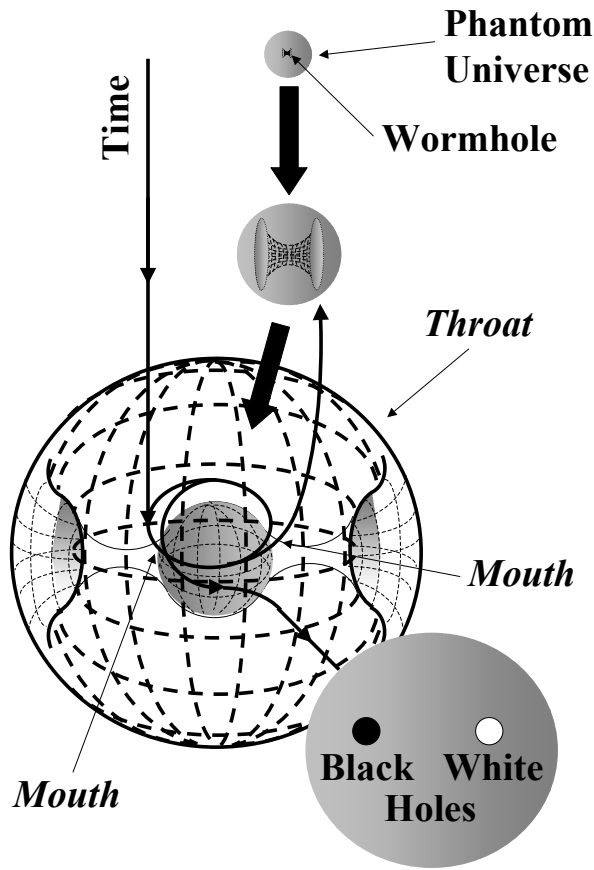}
\caption{\label{fig:epsart} Pictorial representation of the
evolution of a universe filled with phantom energy which is being
accreted onto an initially sub-microscopic wormhole for the case
(B) (see the text) when the wormhole keeps its connections to the
original phantom universe, without resorting to connections to
extra large universes, during the time period on which the
wormhole throat is larger than the universe and time follows
closed curves.}
\end{figure}

The most bizarre implication stemming from the above results is
that for a given cosmic time interval in the future starting at
$t=\bar{t}_-$ and ending at $t=\bar{t}_+$ (for which times
$a=b_0$), first the size of the wormhole throat will exceed the
size of the whole universe, and then, after time $\tilde{t}$, the
universe is contained for a cosmic while inside a giant black hole
(see Figs. 1 and 2). Since, starting at time $\tilde{t}$, whereas
the size of the universe continues steadily increasing toward
infinity, the size of the black hole will gradually decrease
toward zero, the universe will tend to first equalize and then
exceed the black hole size before reaching the big rip. In the
above simpler k-essence model we have
\begin{equation}
\bar{t}_{\pm}= t_b \pm\frac{a_0(t_b\dot{b}_{0i}+
b_{0i})}{2t_b\dot{b}_{0i}b_{0i}}
-\frac{\sqrt{a_0^2(t_b\dot{b}_{0i}+ b_{0i})^2 \pm 4t_b^2a_0
\dot{b}_{0i}b_{0i}^2}}{2t_b t_{0i}\dot{b}_{0i}} ,
\end{equation}
with similar, albeit more complicated expressions for the cosmic
model that corresponds to the scale factor (4).

It follows that during the time interval where the universe is
inside the Morris-Thorne wormhole throat one must necessarily
consider that the wormhole ought to either (A) be first
disconnected from the regions of the original phantom universe at
which it was previously connected, to be instead connected during
that interval to very large regions of two extra larger universes
(such as it is shown on Fig. 2), or (B) keep its connections to
the considered phantom universe according to the topology depicted
on Fig. 3. Since the two wormhole mouths should be moving relative
to one another, due to both the accelerating expansion of the
phantom universe and the wormhole in case (B) and just to the
gradual growth of the wormhole in case (A), in both cases, the
phantom universe could be regarded as just being a "time traveler"
through a gigantic Morris-Thorne wormhole, during that time
interval. Thus, the phantom universe could eventually either
travel to its past to repeat the previous evolution process again,
or travel to its future where it will find itself either inside a
decreasing black hole or containing a smaller black hole when
nearer the big rip, or even evolving, without containing any kind
of holes, on the contracting phase after the big rip [11]. Some of
these possible universal time traveling processes are also
depicted in Figs. 2 and 3.

A key question however remains. Even though Morris-Thorne
wormholes and other topological extensions from Misner space [10]
appear to be stable at scales of the order the Planck length [18],
macroscopic holes have been claimed to be unstable to quantum
vacuum fluctuations [5]. We shall argue nevertheless that
wormholes grown up by accreting phantom energy would keep their
essential quantum nature, and hence their submicroscopic
stability, and should be regarded as quantum spacetime constructs
without classical analog. In fact, by itself, a phantom field can
be considered as a quantum entity in the following sense. It is
known [19] that in order to preserve weak energy condition,
$\rho>0$, the phantom scalar field should be Wick rotated so that
e.g $\phi\rightarrow i\Phi$. Now, it can be seen that such a
rotation is equivalent to Wick rotating the time $t$ itself so
that e.g. $t\rightarrow -i\tau$, while preserving the field
unchanged. By instance, for the dark energy model with scale
factor (4), the scalar field $\phi$ can be expressed as [19]
$\phi+\phi_0=\frac{2}{3\sqrt{1+ w}\ell_P}\ln T$, where $\phi_0$ is
the initial value of the scalar field and we have restored the
Planck length $\ell_P$. In the phantom case $w<-1$ Eq. (20) can be
approximated to $\phi+\phi_0\propto i(t-t_0)$, from which one can
in fact deduce that Wick rotating $\phi$ while keeping $t$
unchanged is equivalent to Wick rotating time $t$ while keeping
$\phi$ unchanged, always preserving the weak energy condition.
Now, it is well known that a Euclideanized spacetime metric would
describe a quantum system [20]. It is thus in a way which
parallels the procedure through which the quantum temperature and
entropy of black holes can be derived [21] that the cosmic phantom
fluid, and hence the wormhole, can be shown to be essential
quantum entities. In fact, it has recently been suggested [22]
that the phantom stuff is characterized by a negative temperature
and becomes therefore an essentially quantum fluid without
classical analog.

An important caveat on the conclusions of this work is worth
mentioning. Even though current data appear to support $w<-1$,
they do not seem to favor a constant equation of state [23].
Therefore, it could well be that the equation of state would relax
back into the stable region of $w>-1$ in the future, before the
big rip singularity or even the wormhole reached an infinite size.
However, even in that case, there would still be some tendency for
the wormhole radius to grow, as we have found.

\acknowledgements

\noindent The author thanks Carmen L. Sig\"{u}enza for useful
discussions. This work was supported by MCYT under Research
Project No. BMF2002-03758.

\end{document}